\documentclass[11pt]{article}
\usepackage[tbtags]{amsmath}
\usepackage{amsfonts}
\usepackage{amssymb}
\usepackage{subfigure}
\usepackage{color}
\usepackage{epsfig}

\newtheorem{thm}{Theorem}
\newtheorem{prop}[thm]{Proposition}
\newtheorem{lem}[thm]{Lemma}

\newcommand{\BlackBox}{\rule{1.5ex}{1.5ex}}
\newcommand{\pf}{\textnormal{Proof: }}
\newcommand{\epf}{\hfill $\BlackBox$}

\newcommand{\beq}{\begin{equation}}
\newcommand{\eeq}{\end{equation}}

\newcommand{\blem}{\begin{lem}}
\newcommand{\elem}{\end{lem}}
\newcommand{\bthm}{\begin{thm}}
\newcommand{\ethm}{\end{thm}}
\newcommand{\bea}{\begin{eqnarray}}
\newcommand{\eea}{\end{eqnarray}}
\newcommand{\bean}{\begin{eqnarray*}}
\newcommand{\eean}{\end{eqnarray*}}

\newcommand{\bit}{\begin{itemize}}
\newcommand{\eit}{\end{itemize}}
\newcommand{\ben}{\begin{enumerate}}
\newcommand{\een}{\end{enumerate}}

\newcommand{\bprop}{\begin{prop}}
\newcommand{\eprop}{\end{prop}}

\title{Diversity Multiplexing Tradeoff of Asynchronous Cooperative Relay Networks}

\author{R. N. Krishnakumar, N. Naveen and P. Vijay Kumar $^{1}$\\
Department of Electrical Communication Engineering \\
Indian Institute of Science \\ Bangalore, India \\
Email: \text{\{kkrn, nnaveen, vijay\}@ece.iisc.ernet.in} }

\begin{document}

\maketitle \thispagestyle{empty} \footnotetext[1]{This work was
carried out while P. Vijay Kumar was at the Indian Institute of
Science, Bangalore, on leave of absence from the Department of
EE-Systems, University of Southern California, Los Angeles, CA 90089
USA (email: vijayk@usc.edu).}

\begin{abstract}
The assumption of nodes in a cooperative communication relay network
operating in synchronous fashion is often unrealistic. In the
present paper we consider two different models of asynchronous
operation in cooperative-diversity networks experiencing slow fading
and examine the corresponding diversity-multiplexing tradeoffs
(DMT). For both models, we propose protocols and distributed
space-time codes that asymptotically achieve the transmit diversity
bound for all multiplexing gains and for any number of relays.
\end{abstract}

\newpage
\tableofcontents
\newpage

\section{Introduction \label{sec:introduction}}

    In fading relay channels cooperative diversity has been introduced
as a technique to provide spatial diversity to help combat fading.
Cooperation creates a virtual transmit antenna array between the
source and the destination that provides the needed spatial
diversity. Cooperative diversity protocols can be broadly classified
as belonging to either the class of Amplify and Forward (AF) or
Decode and Forward (DF) protocols, depending on the mode of
operation of the intermediate relays. In order to fully reap the
benefits of user cooperative diversity, it is necessary for the
network to operate synchronously.  However, in practical distributed
wireless systems, this may be difficult to achieve. As a result,
cooperative-diversity schemes that are designed assuming perfect
timing synchronization, may not be able to fully exploit the
benefits of cooperation in the absence of synchronization. This
motivates the study of cooperation schemes that are robust to
network timing errors.

\begin{figure}[h]
\begin{center}
\includegraphics[height=70mm]{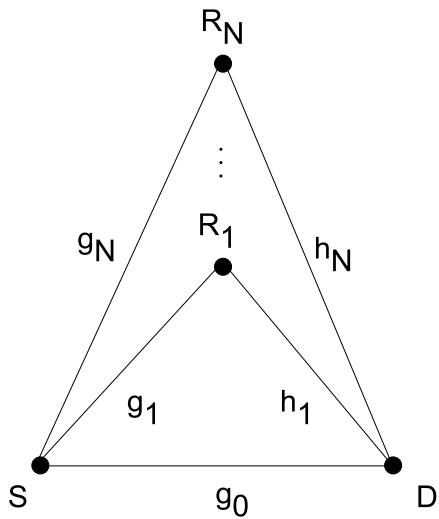}
\caption{Cooperative Relaying in networks. \label{fig:relays_nw}}
\end{center}
\end{figure}

\subsection{Setting and Channel Model \label{subsec:setting and channel}}

We consider two-hop networks with $N$ relays. We will use the
diversity multiplexing tradeoff (DMT) proposed by Zheng and Tse
\cite{ZheTse} as a performance measure. We analyze two-hop networks
with and without a direct link between source and sink. When there
is a direct link, we will assume that the relays are isolated. This
is to make the DMT analysis tractable. In the synchronous case, and
in the presence of relay isolation, there exist protocols that are
known to achieve the optimal DMT of the network,  whereas in the
absence of relay isolation, the optimal DMT of the two-hop network
with direct link remains an open problem. For networks with no
direct link we relax this assumption and are able to handle
scenarios in which there is interference between the relays.

We follow the literature in making the assumptions listed below
concerning the channel. Our descriptions are in terms of the
equivalent complex-baseband, discrete-time channel.\ben \item All
nodes have a single transmit and single receive antenna. \item The
nodes operate in a half-duplex fashion; i.e., at any instant a node
can either transmit or receive but not do both. \item All channels
are assumed to be quasi-static and to experience Rayleigh fading and
hence all fade coefficients are assumed to be i.i.d.,
circularly-symmetric, complex gaussian $\mathbb{C}\mathcal {N}
(0,1)$ random variables.
\item The additive noise at each receiver is also modeled as
possessing an i.i.d., circularly-symmetric, complex gaussian,
$\mathbb{C}\mathcal {N} (0,1)$ distribution. \item The destination
(but none of the relays) is assumed to have perfect knowledge of all
the channel gains. \een

We denote the $i^{th}$ relay by $R_i$. The channel gain from the
source to the relay $R_i$ will be denoted by $g_i$, the gain from
the relay $R_i$ to the destination by $h_i$ and between the relays
$R_i$ and $R_j$ relays by ${\gamma}_{ij}$. The gain for the direct
link, if present, will be denoted by $g_{0}$. All protocols
considered in this work are slotted protocols. This means that the
nodes operate according to a schedule which determines the time
slots in which a node should listen and the time slots in which it
should transmit. In its designated time slot, a node either receives
or transmits a vector of length $T$ channel uses which will be
referred to as a packet.

\subsection{Prior Work \label{subsec:Pri}}

The concept of user cooperative diversity was introduced in
\cite{SenErzAazOne,SenErzAazTwo}. Cooperative diversity protocols
were first discussed in \cite{LanWor} for the two-hop single-relay
network. Zheng and Tse \cite{ZheTse} proposed the
Diversity-Multiplexing gain Tradeoff (DMT) as a tool to evaluate
point-to-point multiple-antenna schemes in the context of slow
fading channels. The DMT was subsequently used as a tool to compare
various protocols for half-duplex, two-hop cooperative networks in
\cite{LanTseWor,AzaGamSch}, see also \cite{YanBelSAF}-\cite{PraVar}.
Reference \cite{SreBirKum} addresses the DMT of the more general
class of multi-hop cooperative relay networks. Jing and Hassibi
(\cite{JinHas}) consider a two hop network without the direct link,
analyze the probability of error and propose code design criteria in
order to achieve full cooperative diversity. All the above works
assume that all relay nodes are perfectly synchronized.

We summarize below results in the recent literature that address
timing errors in cooperative networks.

In \cite{bib:LiXia} and \cite{bib:ShaXia}, the authors analyze
two-hop networks without a direct link and construct distributed
space-time trellis codes that achieve full cooperative diversity
under asynchronism. They consider decode-and-forward two-phase
protocols. Relative propagation delays between source and the relays
as well as from the relays to the destination are assumed to be the
source of timing error. It is assumed that the destination transmits
a beacon to signify reception of a codeword matrix.  In
\cite{bib:ZLiXia}, an Alamouti-based strategy that facilitates
single-symbol decodability is proposed for asynchronous relay
networks, that achieves a diversity order of two for any number of
relays. Codes with low decoding complexity achieving full diversity
for any number of relays, in the presence of asynchronism were
constructed in \cite{SusRaj}.

Damen and Hammons (\cite{bib:DamHam}) construct delay-tolerant
distributed TAST block codes that achieve full diversity for the
two-phase protocol in the presence of delays, under the decode and
forward (DF) strategy. Their constructions are flexible in terms of
accommodating a varying number of transmit and receive antennas as
well as varying decoding complexity.

Wei considers the two-hop network with delays in \cite{bib:Wei} and
analyzes the DMT of certain protocols. For the two phase DF
protocol, \cite{bib:Wei} considers both the scenario where the relay
performs independent coding as well as one in which joint
distributed space time coding is carried out and the author then
goes on to evaluate the DMT of the two schemes. However, these
schemes do not meet the cut-set bound on DMT for all values of
multiplexing gain.

\subsection{Results}
We consider two different models of asynchronous operation in
cooperative relay networks: \ben \item the propagation-delay model
\item the slot-offset model. \een For both these models we propose
a variant of the Slotted Amplify and Forward (SAF) protocol proposed
in \cite{YanBelSAF} which asymptotically achieves the transmit
diversity bound in the absence of a direct source-destination link,
for any number of relays and for an arbitrary delay profile. When
there is a direct link we achieve the transmit diversity bound under
the assumption that all the relays are isolated. We also present DMT
optimal codes for both cases considered above.

\section{Slotted Amplify and Forward Protocol for the Propagation-Delay Model \label{sec:SAFprop}}

\subsection{Propagation Delay Model}

Under this model we take into account the relative propagation
delays between the various nodes. We assume that the delay is in
units of one symbol duration (i.e., one channel use). We adopt the
following notation: the pair $(\nu_{i}, \pi_{i})$ denotes the delay
between source and relay $R_i$ and between relay $R_i$ and the
destination respectively. The overall delay in the $i^{th}$ relay
path will be denoted by $\tau_{i}$. Thus $\tau_{i} = \nu_{i} +
\pi_{i}$. Set \[ \theta \ = \ \max_{1 \leq i \leq N} {\tau_i}.
\]
We will make the simplifying assumption that the delay between any
two relays is zero.  It turns out that the lower bounds on DMT
that we derive here, will only improve in the presence of
inter-relay delays. All delays are assumed to be known at the
destination. The relays are assumed to know the propagation delay
from source to the respective relay to the extent that that the
relay node knows when it is receiving its intended packet from the
source.

Synchronous network operation implies in particular, packet-level
synchronization.  As a result, packets arriving at either a relay
node or the destination, originating at either the source or a relay
node, are aligned in time  and thus it is meaningful to speak of
interfering packets. This will no longer be the case if there exists
relative propagation delays as there will now be relative
time-shifts between packets arriving from various nodes, either at a
relay node or at the destination.

\subsection{DMT Analysis of the Quasi-Synchronous Network under the Naive SAF
Protocol \label{subsec:DMT analysis of the quasi-synchronous
network}}

\begin{figure}[h]
\begin{center}
\includegraphics[height=70mm]{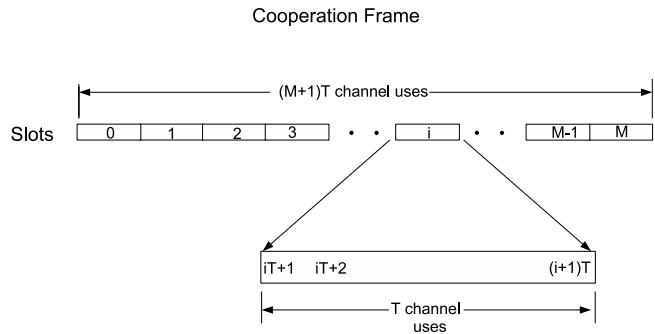}
\caption{Cooperation frame for SAF. \label{fig:SAFframe}}
\end{center}
\end{figure}

\begin{figure}[h]
\begin{center}
\includegraphics[width=90mm]{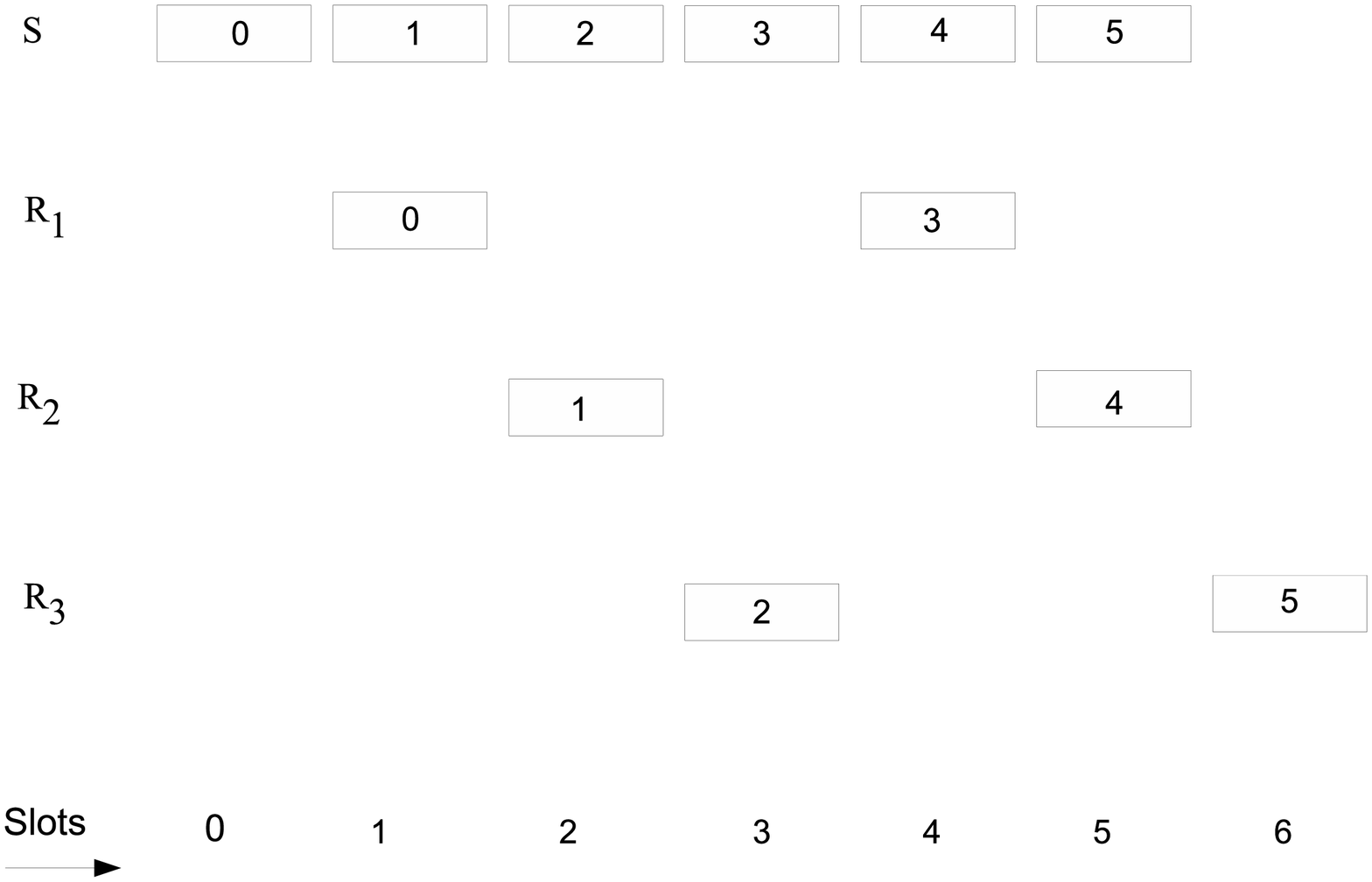}
\caption{SAF protocol for the case of $N = 3$ relays, and a total of
$M+1 = 7$ slots. \label{fig:SAFslots}}
\end{center}
\end{figure}

We consider a particular AF protocol, known in the literature as the
Slotted Amplify and Forward (SAF) protocol, and proposed by Yang and
Belfiore~\cite{YanBelSAF}. We give a brief description of the
protocol here as applied to the case when there is no direct link
from source to destination. An $N$-relay $M$-slot SAF operates as
follows: \ben \item The protocol splits up the time axis into frames
and slots.  There are $T$ channel uses per slot and $(M+1)$ slots
per frame, where $M=KN$ is a multiple of the number $N$ of relays.
The slots are indexed from $0$ through $M$. If one ignores the first
initialization slot (slot zero), then there are in effect $M$ slots
per frame.  Under the SAF protocol, the $M$ slots are divided into
$K$ cycles, each cycle being of time duration equal to $NT$ channel
uses (see Fig.~\ref{fig:SAFframe}).\item In the first,
initialization slot, the source transmits $\underline{x}_{0}$ and
during the $i^{th}$ slot $1 \leq i \leq M-1$, the vector
$\underline{x}_i \ = \ [x_{iT + 1} \ x_{iT + 2} \ \ldots \
x_{(i+1)T}]$ is transmitted.
\item  During the $i$th time slot, $1 \leq i \leq M$, the $j$th
relay, $R_j$, where $j$, $ 1 \leq j \leq N$, is such that $j \ = \ i
\pmod{N}$, forwards the $\underline{s}_i$ signal received by it
during the immediately preceding, i.e., $(i-1)$th slot.  However, to
simplify the description, we will write $R_i$ in place of $R_j$.
Thus for example, if $N=4$, during the $7$th time slot, we will
speak of relay $R_7$ as forwarding the signal received by it during
the $6$th time slot, when we actually mean relay $R_3$, since $7=3
\pmod{4}$. Note that $\underline{s}_i$ includes the additive noise
present in the receiver of the $i$th relay. One round of
transmissions by all the relays constitutes a cycle. Thus each frame
is comprised of $K$ cycles with each relay transmitting precisely
once during each cycle. The source remains silent in the final,
$M^{th}$ slot (see Fig.~\ref{fig:SAFslots}). \een
We now write down an expression for the signals received by relay
and destination during the $i^{th}$ slot.  The relay $R_{i+1}$
receives \bea \underline{s}_{i} &=& g_{i+1}\underline{x}_{i} +
\gamma_{i,i+1}\underline{s}_{i-1} + \underline{v}_{i}
\label{eq:SAF_signal_at_relay}\eea where $\underline{s}_{i-1}$ is
the signal transmitted by $R_i$. The signal received at the
destination is given by \bea \underline{y}_i =
h_i\underline{s}_{i-1} + \underline{w}_{i}.\eea Here
$\underline{v}_{i}$ and $\underline{w}_{i}$ denote white noise.
Fig.~\ref{fig:SAFslots} depicts the protocol operation for $N = 3$
and $M = 6$.
The destination collects the vector of received symbols and at the
end of one frame the resulting channel model can be written as,
\[
\underline{y} = H\underline{x} + G\underline{v} + \underline{w}.
\]
Here $\underline{x} = \left[\underline{x}_0 \ \underline{x}_1 \
\ldots \ \underline{x}_{M-1}\right]$ is the vector of transmitted
symbols and $H$ is a lower triangular matrix of size $MT \times MT$.
The main diagonal of $H$ consists of entries of the form $\gamma_i =
h_ig_i$, each entry being repeated $K$ times. $G$ is a matrix
consisting of channel gains. The resulting noise term
$G\underline{v} + \underline{w}$ is not white in general. But it has
been shown in \cite{SreBirKum} that for the class of AF protocols in
general the resulting noise at the destination is white in the scale
of interest. Hence from now on we will work with the model
\[
\underline{y} = H\underline{x} + \underline{w}
\] with $\underline{w}$ being white. The DMT of
the above matrix can be lower bounded by
\[
d(r) \geq  N\left(1 - \frac{M + 1}{M}r\right)
\] which meets the transmit diversity bound as $M$ tends to
infinity. For the case with direct link, the protocol remains the
same as above except that the source transmits in the $M^{th}$ slot
as well. This is shown to asymptotically achieve the DMT $(N + 1)(1
- r)$ under the assumption that all the relays are isolated; i.e., a
relay cannot hear what the other relays transmit( see
\cite{YanBelSAF}). However the exact DMT of the SAF protocol without
the assumption of relay isolation remains an open problem. A lower
bound to the DMT of the SAF protocol under the assumption of
relay-isolation can also be found derived in \cite{SreBirKum}.

\subsubsection{Impact of Propagation Delays on the DMT}\label{subsubsec:Impact of delays}

We now analyze the impact of delays on the operation of the naive
SAF protocol for the case of a two-hop network without direct link.
Throughout this sub-section, we will assume quasi-synchronous
operation of the network, by which we will mean that there is no
propagation delay between the source and any of the relay nodes.
Thus the value $\nu_i, i=1,2, \ldots, N$ of these delays from the
source to the relays are all set equal to zero. In order to isolate
different frames, the source extends the silence in the last slot to
$T+\theta$ channel uses in place of $T$ channel uses in naive SAF.
The analysis of this simple network will serve to illustrate the
performance degradation of naive SAF and will also provide insights
on how to handle the general case when both $\nu_i$ and $\pi_i$ are
non zero.  We now introduce some terminology. When the symbols in a
packet are arranged chronologically, those appearing at the
beginning will be called the head of the packet while the tail of
the packet will correspond to symbols at the end.

\begin{figure}[h]
 \centering
  \subfigure[Reception at the relays]{\label{fig:relayreception}\includegraphics[height=50mm]{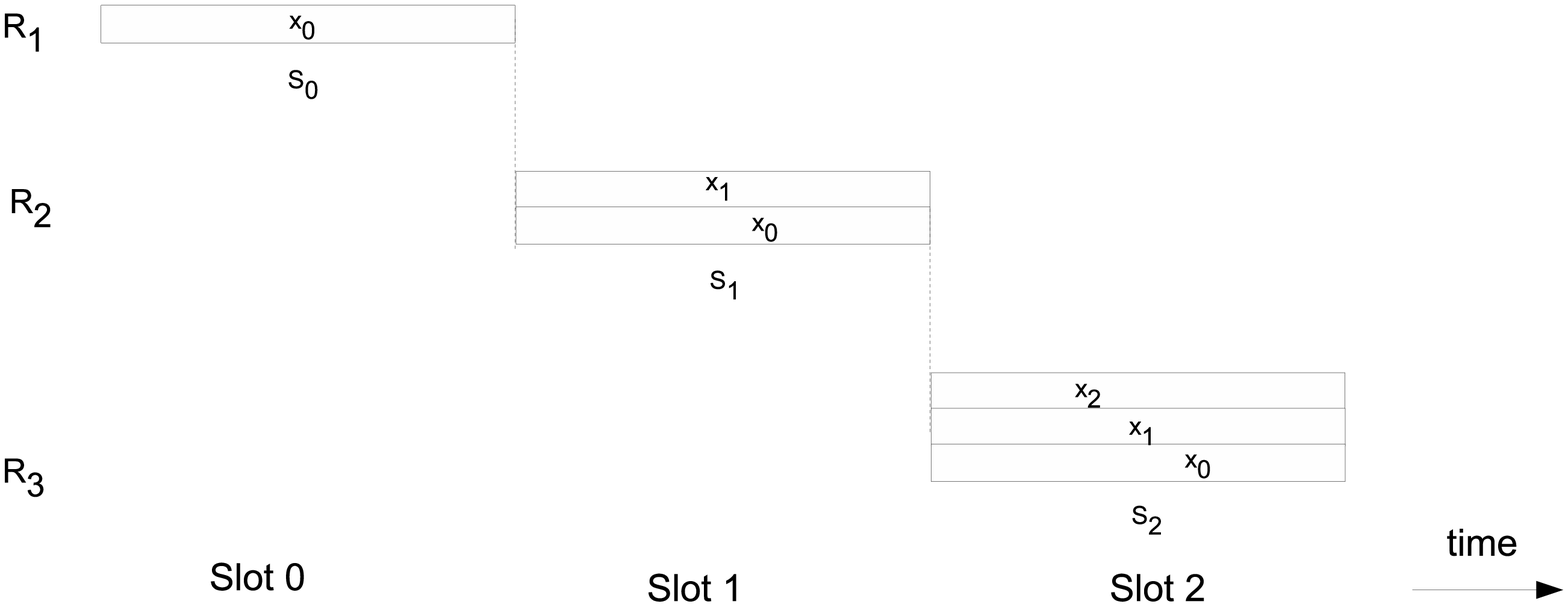}}
  \subfigure[Reception at the destination]{\label{fig:destreception}\includegraphics[height=50mm]{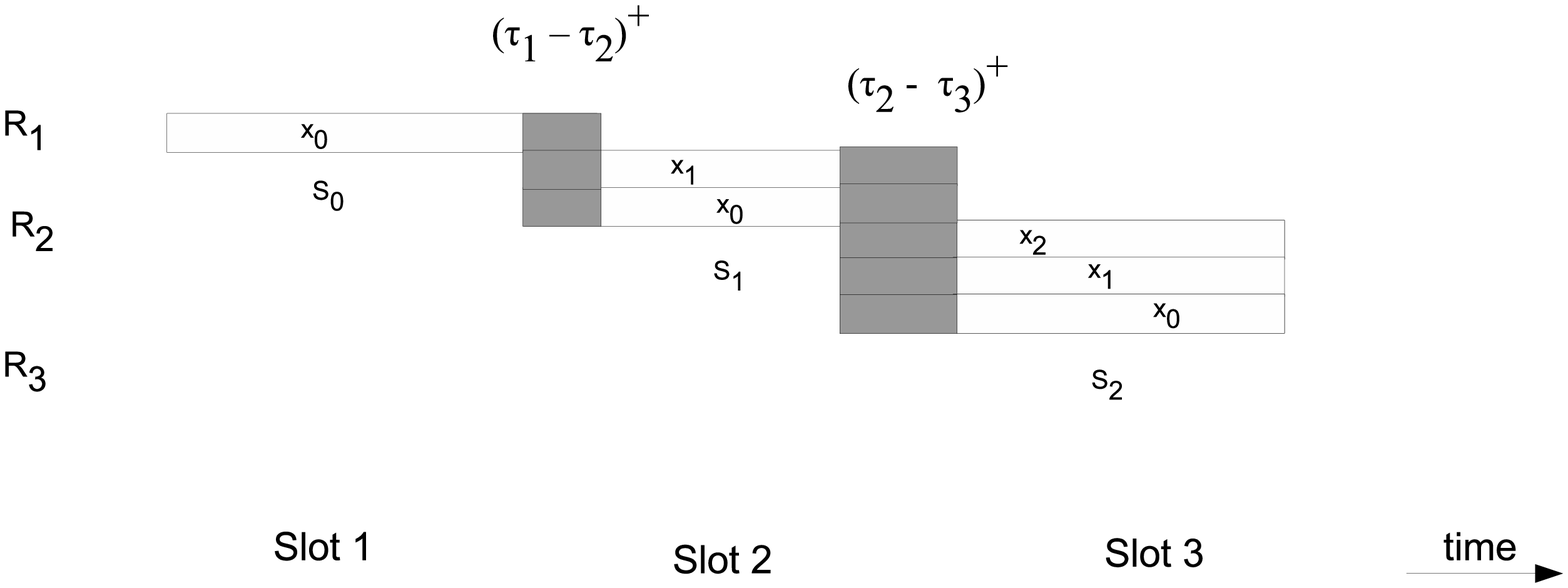}}
  \caption{Reception at the relays and destination}
  \label{fig:relaydestrx}
\end{figure}

Since each $\nu_i=0$, the signals received at the relays are exactly
the same as that in the case of perfect synchronization (see
Fig.~\ref{fig:relayreception}). As per the SAF protocol, while the
relay $R_{i+1}$ is listening to packet $\underline{x}_i$ from the
source in the $i^{th}$ slot, the relay $R_{i}$ will be transmitting
packet $\underline{s}_{i-1}$. Thus relay $R_{i+1}$ will
simultaneously receive transmissions from the source and relay node
$R_{i}$ and these transmissions will moreover, be aligned in time.
This however, is not the situation at the destination. Depending
upon the value of the relative delays $\pi_i$, the tail of the
transmission from relay node $R_{i}$ could for instance, interfere
with the head of the transmission from relay node $R_{i+1}$. This is
depicted in Fig.~\ref{fig:destreception}.
The receptions at the destination will get modified as, \bea
\underline{y}_i &=& h_i \underline{s}_{i-1} + h_{i-1}D^{T - \Delta
\tau_{i-1}}(\underline{s}_{i-2}) + h_{i+1}D^{T -\Delta
\tau_{i}}(\underline{s}_{i}) + \underline{w}_{i}.\eea

Notice that in addition to the intended reception from relay $R_i$
there is interference from relays $R_{i-1}$  and $R_{i+1}$. Here
$D^{T -\Delta \tau_{i}}$ corresponds to a shift-and-truncate
operator acting on a vector of symbols. The direction of shift
depends upon the sign of $\Delta \tau_{i} = (\tau_i - \tau_{i+1})$.
If $\Delta \tau_i$ is positive (negative) then $D^{T -\Delta
\tau_{i}}\underline{x}$ shifts $\underline{x}$ to the right (left)
by $(T -|\Delta \tau_{i}|)$ symbols and drops the last (first) $(T
-\Delta \tau_{i})$ symbols.

The signal received by the destination over the course of one frame
can be expressed in the form
\[ \underline{y} = H\underline{x}  + \underline{w}.
\]Here, unlike in the synchronous case, the resulting channel matrix $H$ will no longer be
lower-triangular. We illustrate with an example where $N = 3, M = 3,
T = 3$.   In the case of perfect synchronization, the channel matrix
at the destination would look like, \bea \left[
\begin{array}{c}
        y_0\\
        y_1 \\
        y_2 \\
        y_3 \\
        y_4\\
        y_5 \\
        y_6 \\
        y_7 \\
        y_8 \\
        \end{array}
        \right] &=& \left[
\begin{array}{ccccccccc}
        \gamma_1\\
        & \gamma_1 \\
        && \gamma_1 \\
        \beta_1&&& \gamma_2 \\
        &\beta_1&&& \gamma_2 \\
        &&\beta_1&&& \gamma_2 \\
        \beta_3&&&\beta_2&&&\gamma_3\\
        &\beta_3&&&\beta_2&&&\gamma_3\\
        &&\beta_3&&&\beta_2&&&\gamma_3\\
        \end{array}
        \right] \left[
\begin{array}{c}
        x_0\\
        x_1 \\
        x_2 \\
        x_3 \\
        x_4\\
        x_5 \\
        x_6 \\
        x_7 \\
        x_8 \\
        \end{array}
        \right] + \underline{w} \notag \eea where $\beta_1 = g_1h_2\gamma_{12}$, $\beta_2 =
g_2h_3\gamma_{23}$, $\beta_3 = g_1h_3\gamma_{12}\gamma_{23}$.
However, in the presence of delays $\pi_1 = 2, \ \pi_2 = 1$  and $
\pi_3 = 0$, the resulting channel matrix takes on the form \bea
\left[
\begin{array}{c}
        y_0\\
        y_1 \\
        y_2 \\
        y_3 \\
        y_4\\
        y_5 \\
        y_6 \\
        \end{array}
        \right] &=& \left[
\begin{array}{ccccccccc}
        \gamma_1\\
        & \gamma_1 \\
        \beta_1&&\gamma_1&\gamma_2 \\
        &\beta_1&&& \gamma_2 \\
        \beta_3&&\beta_1&\beta_2&& \gamma_2&\gamma_3 \\
        &\beta_3&&&\beta_2&& &\gamma_3 \\
        &&\beta_3&&&\beta_2&& &\gamma_3 \\
        \end{array}
        \right] \left[
\begin{array}{c}
        x_0\\
        x_1 \\
        x_2 \\
        x_3 \\
        x_4\\
        x_5 \\
        x_6 \\
        x_7 \\
        x_8 \\
        \end{array}
        \right] + \underline{w}. \label{eq:channel_eg_with_delays}
\eea  It is not easy, in general, to compute the DMT of channel
matrices $H$ having structure of the form appearing above. The
difficulty arises precisely from terms like $\gamma_{i}x_i +
\gamma_{i+1}x_i$ that are heard at the destination due to
simultaneous reception from two relays. One possible way to lower
bound the DMT is to drop some symbols from the received vector such
that the channel matrix between the remaining subset of input and
output symbols is lower triangular. For if the matrix is lower
triangular, we can use results from \cite{SreBirKum} to lower bound
the DMT. For the channel in equation
\ref{eq:channel_eg_with_delays}, it is clear that after dropping
$\{y_2, y_4, y_6\}$ the resulting channel model will be \bea \left[
\begin{array}{c}
        y_0\\
        y_1 \\
        y_3 \\
        y_5 \\
        \end{array}
        \right] & = & \left[
\begin{array}{cccc}
        \gamma_1\\
        & \gamma_1 \\
        &\beta_1&\gamma_2 \\
        &\beta_3&\beta_2&\gamma_3 \\
        \end{array}
        \right] \left[
\begin{array}{c}
        x_0\\
        x_1 \\
        x_4 \\
        x_7 \\
        \end{array}
        \right]. \label{eq:channel_eg_lowertriangular}
\eea We now show through a sequence of mutual information
inequalities that the DMT of the resulting channel, after dropping
certain terms, is indeed a lower bound for the DMT of the original
matrix. Let $\underline{x}_A$ and $\underline{y}_B$ be the vector of
input and output symbols that we retain. Then \bea
I(\underline{x};\underline{y}|H) &=& I(\underline{x}_A,
\underline{x}_{A^c};\underline{y}_B ,\underline{y}_{B^c}|H) \notag \\
&\geq& I(\underline{x}_A,\underline{x}_{A^c};\underline{y}_B|H) \notag \\
&=& I(\underline{x}_A;\underline{y}_B|H) + I(\underline{x}_{A^c};\underline{y}_B|\underline{x}_{A},H )\notag \\
&\geq& I(\underline{x}_A;\underline{y}_B|H) =
I(\underline{x}_A;\underline{y}_B|H') \eea where $H'$ is the channel
relating $\underline{x}_A$ and $\underline{y}_B$, i.e.,
\[\underline{y}_B = H'\underline{x}_A + \underline{w}'.\]

We now provide a systematic way in which we can drop symbols at the
destination so that the resulting channel matrix is lower
triangular. This is by identifying and removing unintended
interference at the destination. We call the interference between
the tail of a packet and the head of the next immediate packet at
the destination as collision. It is clear that since collision is an
event that does not happen in the synchronous case, we have to drop
the symbols involved in collision in order to retain the channel
structure. Now, say packet $\underline{s}_{i-2}$ collided with
$\underline{s}_{i-1}$ but $\underline{s}_{i-1}$ did not collide with
$\underline{s}_{i}$. From equation (\ref{eq:SAF_signal_at_relay}) it
is clear that the symbols in $\underline{s}_{i-1}$ that were
involved in collision would be present in the form of interference
in $\underline{s}_{i}$. This interference is also unintended and
hence these symbols have to be dropped when $\underline{s}_{i}$
reaches the destination. We call such interference as corruption. It
is different from collision since it happens at the relays. For if
the relays were isolated there would be no corruption at all. It
must be noted that collision is the cause of corruption. In a
network where all the source to destination delays are equal but
arbitrarily split between the source to relay and the relay to
destination paths, the DMT of the SAF protocol remains unchanged.

Let us bound the number of symbols that need to be dropped from
every packet due to collisions and corruptions. We may drop symbols
from packet $\underline{y}_{i}$ as a consequence of one or more of
the following events:\ben
\item Collision at head with $\underline{s}_{i-2}$.
\item Collision at tail with $\underline{s}_{i}$.
\item Corruption at tail due to $\underline{s}_{i-2}$.
\een

The number of symbols involved in the collision at head is
$(\pi_{i-1} - \pi_{i})^{+}$ and similarly at tail it is $(\pi_{i} -
\pi_{i+1})^{+}$. The number of corrupted symbols in the tail of
$\underline{s}_{i-1}$ is again at most $(\pi_{j} - \pi_{j+1})^{+}$
for some $j$. Hence the number of symbols that need to be dropped
can be bounded by $2\max_j(\pi_j - \pi_{j+1})^+$. Since we are
interested in designing protocols for a certain maximum delay
$\theta$ we further bound this as $2\theta$. Thus if we drop
$T-2\theta$ symbols from each slot of the received vector, the
resulting channel matrix will be lower triangular, the DMT of which
can be bounded easily.

\subsubsection{Lower Bound on DMT}\label{subsubsec:LB on DMT of SAF for quasi synchronous
network}

At the end of a frame, after dropping some symbols, the destination
would have received a vector of length $M(T - 2\theta)$. The channel
model can be written as, \bea \underline{y} & = & H\underline{x}^{}
+ \underline{w} \label{eq:Channel_for_naive_SAF}\eea where
 \bea
    {\underline{x}}^t & = & \left[\begin{array}{ccccc}
                    \underline{x_0}^{t} & \underline{x_1}^{t} & \underline{x_2}^{t} &
                    \cdots & \underline{x_{M-1}}^{t} \end{array} \right]
\eea is the vector of clean symbols, $H$ is the $M(T-2\theta)\times
M(T-2\theta)$ channel matrix and $\underline{w} \  \epsilon \
\mathcal{C}^{M(T-2\theta)}$ is the noise vector. The channel matrix
$H$ is lower triangular. The structure of the lower triangular half
of the matrix depends on the delay profile of the network. We have
the following result.

\bprop \label{prop:DMT delay model no DL} For the protocol described
in section \ref{subsec:DMT analysis of the quasi-synchronous
network} the DMT is lower bounded by
\[
d(r) \geq  N\left(1 - \frac{(M+1)T + \theta}{M(T-2\theta)}r\right)
\] which asymptotically approaches the transmit diversity bound for
large $M$ and $T$. \eprop

    In order to derive a lower bound on the DMT of the proposed protocol
we make use of the following important result from \cite{SreBirKum}.

\bthm \label{thm:LB on DMT}(\cite{SreBirKum}) Let $H$ be a lower
triangular matrix with random entries. Let $H_d$ be a diagonal
matrix whose diagonal entries are the same as the main diagonal
entries of $H$, and let $H_l$ be a matrix consisting of only the
last sub diagonal of $H$ and zeros else where. Let $d_H(r),
d_{H_d}(r)$ and $d_{H_l}(r)$ denote the DMT of these matrices
respectively. Then \ben
\item $d_H(r) \geq d_{H_d}(r)$.
\item $d_H(r) \geq d_{H_l}(r)$.
\item In addition, if the entries of $H_d$ and $H_l$
are independent, then $ d_H(r) \geq d_{H_d}(r) + d_{H_l}(r) $. \een
\ethm

{\bf Proof of proposition \ref{prop:DMT delay model no DL}:}
Consider the channel model given by
(\ref{eq:Channel_for_naive_SAF}). Let $T^{'} = T - 2\theta$. Let
$H_d$ be the corresponding diagonal matrix associated with $H$. The
diagonal entries of $H_d$ are the product fade coefficients $h_ig_i$
each repeated $KT^{'}$ times. Let us denote $\gamma_i = h_ig_i $.
The entries below the principal diagonal are products of $h_i$'s and
$g_i$'s which are correlated with $\gamma_i$'s. Hence we make use of
implication 1 of theorem \ref{thm:LB on DMT}. Let $ r^{'} = ((M+1)T
+ \theta)r$. The DMT of the diagonal matrix $H_d$ can be easily
calculated as follows: \bea
Pr\{I(\underline{x};\underline{y}|H_d)\leq r^{'}\log\rho\} & = &
Pr\{\log\det(I + \rho H_d H_d^{\dagger}) \leq r^{'}\log\rho\}\\
& = & Pr\{\log(\prod_{i=1}^{N}(1 + \rho|\gamma_i|^2)^{KT^{'}}) \leq
r^{'}\log\rho\}\\
& = & Pr\{KT^{'} \sum_{i=1}^{N}(1-\alpha_i)^{+} \leq r^{'}\} \eea
where $|\gamma_i|^2 = \rho^{-\alpha_i}, i = 1,2,\ldots,N$. Therefore
we have, \bea
 d_{H_d}(r) & = & \inf_{KT^{'} \sum_{i=1}^{N}(1-\alpha_i)^{+}
\leq r^{'}}(\sum_{i=1}^{N}\alpha_i)\\
& = & N\left(1 - \frac{(M+1)T + \theta}{M(T-2\theta)}r\right). \eea
By theorem \ref{thm:LB on DMT}, a lower bound on the DMT of the
proposed protocol can be given by
\[
d_H(r) \geq  d_{H_d}(r) = N\left(1 - \frac{(M+1)T +
\theta}{M(T-2\theta)}r\right)
\] which asymptotically approaches the transmit diversity bound for
large $M$ and $T$. As M tends to infinity we have \bea\label{eqn:DMT
of naive SAF under delay with N alone going to infinity} d_H(r) \geq
N\left(1 - \frac{T}{T-2\theta}r\right). \eea \epf

From equation (\ref{eqn:DMT of naive SAF under delay with N alone
going to infinity}), we see that there is a rate loss factor of
$\frac{T}{T-2\theta}$. This factor reflects the loss in maximum
multiplexing gain for a finite slot length but arbitrarily large
number of cycles. This is actually the ratio of the length of one
slot to the number of clean symbols per slot for the protocol. In
the next sub-section we show how padding zeros at the source helps
in increasing the number of clean symbols per slot, thereby
increasing the resulting DMT.

\subsection{SAF Protocol with Guard Time for the Network with Arbitrary Delay Profile\label{subsec:SAF guardtime propagation delay}}

From the analysis in section \ref{subsubsec:Impact of delays}, it is
clear that a simple way to ensure lower-triangular structure of the
channel matrix is to include a guard time for each slot, thereby
avoiding collisions at the destination. By guard time we mean a
period of silence where the source transmits nothing, thereby
slowing down its operation. The guard time ensures that the number
of colliding symbols at the destination is reduced.

Let $x$ be the length of the guard time per slot. Now the effective
length of every slot is $T^{'} = T + x$. The strategy adopted by the
relays is as follows: Every relay listens to the source for $T$
channel uses and transmits in the next $T$ immediate channel uses.
As described in the previous sub-section the destination collects
only the collision free symbols in every slot. Then it is easy to
see that the number of clean symbols in every slot can be lower
bounded by $(T - 2\theta + 2x)$. Thus adding $x$ guard symbols
provides a two-fold increase in the number of clean symbols.
Following similar outage analysis as in section
\ref{subsubsec:Impact of delays}, the DMT of the resulting SAF
protocol with guard time can be lower bounded as, \bea d(r) & \geq &
N \left(1 - \frac{T + x}{T - 2\theta + 2x}r\right) \label{eq:LB with
guard time}. \eea In particular when $x = \theta$ we have \bea d(r)
& \geq & N \left(1 - \frac{T + \theta}{T}r\right). \label{eq:LB with
guard time1} \eea The lower bound in equation (\ref{eq:LB with guard
time1}) is better than that in equation (\ref{eq:LB with guard
time}) for $x \leq \theta$. Thus padding $\theta$ zeros in every
slot gives the optimal guard time and it completely eliminates
collision. Padding more than $\theta$ zeros only reduces the DMT as
the number of clean symbols in every slot can at most be only $T$.

Even for the more general case where there are delays from the
source to the relays as well, it can be shown through a similar
analysis as in section \ref{subsec:DMT analysis of the
quasi-synchronous network} that the number of clean symbols per slot
for naive SAF can be bounded as ($T - 2(\tau_{\max} - \tau_{\min})$)
in place of ($T - 2\max\{(\tau_i - \tau_{i+1})\}$) for the quasi
synchronous network. A guard time of $\theta$ would be sufficient in
this case, the details of which are provided in the next section.

\subsubsection{Case with no Direct Link \label{subsubsec:SAF guardtime no direct link}}

We first consider the two-hop network without the direct link. We
show that the SAF protocol with guard time asymptotically achieves
the transmit diversity bound for any delay profile (${\nu}_i,
{\pi}_i$), $i = 1,2,\ldots, N$. The description of the protocol is
as follows: \ben
\item There are totally $(M+1)$ slots in a frame. The source operation is the
same as that in naive SAF described in section \ref{sec:SAFprop},
except that in each of the first $M$ slots the source flushes
$\theta$ zeros after sending $T$ information symbols. The total
duration of one frame is $(M(T + \theta) + T)$ channel uses. \item
Each relay in its respective slot, listens for $T$ channel uses and
transmits in the next immediate $T$ channel uses. \een The signal
received by the relay $R_{i+1}$ in the $i^{th}$ slot is given by,
\bea \underline{s}_{i} &=& g_{i+1}\underline{x}_{i} +
D^{\Delta\nu_i} \ \gamma_{i,i+1}\underline{s}_{i-1} +
\underline{v}_{i} \eea where $\underline{s}_{i-1}$ is the signal
transmitted by relay $R_i$. Here $D^{\Delta\nu_i}$ is again a
shift-and-truncate operator which operates as follows. If $\Delta_i$
is positive (negative) then $D^{\Delta\nu_i}$ shifts $\underline{x}$
to the right (left) by $|\Delta\nu_i|$ symbols and drops the last
(first) $|\Delta\nu_i|$ symbols.

The signal transmitted by the relay $R_i$ arrives at the destination
with a delay of $\pi_{i}$. By padding $\theta$ zeros we ensure that
the receptions at destination from different relay nodes are made
orthogonal. It is clear that these receptions will remain orthogonal
even if there are inter-relay delays. Hence adding a guard time
Notice that the different relay transmissions will also now be
orthogonal. There is a total delay of $\tau_{i}$ in the $i^{th}$
path from the source to the destination. As the destination is
assumed to know all the delays in the network it knows exactly when
the transmission and the reception of a particular relay starts and
ends and hence listens to the network only during these time
instants. At the end of one frame the destination would have
received a vector of length $MT$ consisting of the receptions from
all the relays. The channel model is given by, \bea \underline{y} &
= & H\underline{x} + \underline{w} \label{eq:channel model for
SAF}\eea where
 \bea
    \underline{x}^{t} & = & \left[\begin{array}{ccccc}
                    \underline{x}_{0}^{t} & \underline{x}_{1}^{t} & \underline{x}_{2}^{t} &
                    \cdots & \underline{x}_{M-1}^{t} \end{array} \right]
\eea is the vector of transmitted symbols from the source, $H$ is
the $MT\times MT$ channel matrix and $\underline{w} \  \epsilon \
\mathcal{C}^{MT}$ is the noise vector. The channel matrix $H$ is
lower triangular, which is a result of padding $\theta$ zeros to
every slot. We then have the following theorem.

\bthm \label{thm:DMT delay model no DL} For the protocol described
in section \ref{subsubsec:SAF guardtime no direct link} the DMT is
lower bounded by
\[
d(r) \geq  N\left(1 - \frac{M(T + \theta) + T}{MT}r\right)
\] which asymptotically approaches the transmit diversity bound for
large $M$ and $T$. \ethm

\pf Consider the channel model given by (\ref{eq:channel model for
SAF}). Let $H_d$ be the corresponding diagonal matrix associated
with $H$. The diagonal entries of $H_d$ are the product fade
coefficients $h_ig_i$ each repeated $KT$ times, where $K =
\frac{M}{N}$. Let us denote $\gamma_i = h_ig_i $. Let $ r^{'} = (M(T
+ \theta) + T)r$. The DMT of the diagonal matrix $H_d$ can be easily
calculated as follows: \bea
Pr\{I(\underline{x};\underline{y}|H_d)\leq  r^{'}\log\rho\} & = &
Pr\{\log\det(I + \rho H_d H_d^{\dagger}) \leq r^{'}\log\rho\}\\
& = & Pr\{\log(\prod_{i=1}^{N}(1 + \rho|\gamma_i|^2)^{KT}) \leq
r^{'}\log\rho\}\\
& = & Pr\{KT \sum_{i=1}^{N}(1-\alpha_i)^{+} \leq r^{'}\} \eea where
$|\gamma_i|^2 = \rho^{-\alpha_i}, i = 1,2,\ldots,N$. Therefore we
have,

\bea
 d_{H_d}(r) & = & \inf_{KT \sum_{i=1}^{N}(1-\alpha_i)^{+}
\leq r^{'}}(\sum_{i=1}^{N}\alpha_i)\\
& = & N\left(1 - \frac{M(T + \theta) + T}{MT}r\right). \eea By
theorem \ref{thm:LB on DMT}, a lower bound on the DMT of the
proposed protocol is then given by,
\[
d_H(r) \geq  N\left(1 - \frac{M(T + \theta) + T}{MT}r\right)
\] which asymptotically approaches the transmit diversity bound for
large $M$ and $T$. As the number of slots $M$ tends to $\infty$ we
have
\[
d_H(r) \geq  N\left(1 - \frac{T + \theta}{T}r\right).
\]\epf

The naive SAF protocol achieves $N(1 - r)$ as M goes to infinity in
the case of perfect synchronization, where as in the presence of
asynchronism the lower bound on the DMT of the modified SAF is
$N\left(1 - \frac{T + \theta}{T}r\right)$. Thus we see that the
impairment due to asynchronism in a cooperative relay network can be
combated by operating the protocol over a larger time duration. From
the expression for the lower bound on DMT, we see that there is a
rate loss of $\frac{T}{T + \theta}$ which is negligible for large
$T$.

The DMT of the channel given by (\ref{eq:channel model for SAF}) can
be achieved by using an approximately universal Cyclic division
Algebra (CDA) based code of size $MT \times MT$, where the coding
takes place over $MT$ frames. The details of the code construction
are described in section \ref{sec:DMT optimal codes}.

\subsubsection{Case with Direct Link \label{subsubsec:SAF guardtime direct link}} Here
we consider the two hop network with the direct link. The additional
assumption we make in this section is that all the relay nodes are
isolated from each other, i.e., $\gamma_{ij} = 0,\  \forall \ i,j$.
Let the delay in the direct link be $\tau_0$. Here $\theta = \max_{i
= 0,1,2, \ldots, N}\tau_i$

The strategy adopted by the source and the relays is the same as
that in the previous section, except that the source continues to
transmit in the last $M^{th}$ slot. We take the number of slots to
be $(M + 1)$. The length of one frame is $[(M + 1)(T + \theta)]$
channel uses.

The strategy adopted by the destination is as follows: Since the
destination knows all the delays in the network it collects the
received symbols only when the source is transmitting (it just drops
the instants when the source is idle). For example it collects the
received samples during the following time instants.
\[
\tau_0 + 1 \rightarrow T + \tau_0,\  T + \theta + \tau_0 + 1
\rightarrow 2T + \theta + \tau_0 + 1, \ldots
\]
At the end of one frame the destination would have received a vector
of length $(M + 1)T$. The resulting channel model is given by, \bea
\underline{y} =  H\underline{x} + \underline{w} \label{eq:channle
model for SAF with D} \eea where $H$ is a $(M + 1)T \times (M + 1)T$
matrix.

\bthm \label{thm:LB delay model DL} For the protocol described above
the DMT is lower bounded by \bea
    d(r) & \geq & \left(1 - \frac{(M + 1)(T + \theta)}{(M +
    1)T}r\right) + N\left(1 - \frac{(M + 1)(T + \theta)}{M(T - \theta)}r
    \right)\notag
\eea which asymptotically achieves the transmit diversity bound for
large $M$ and $T$. \ethm

\pf  The lower triangular matrix $H$ consists of just two band of
entries, the main diagonal and the sub diagonal immediately below
the main diagonal. This is a consequence of the assumption that all
the relays are isolated. Next we observe the following: \ben
\item The main diagonal of $H$ contains $g_0$ repeated $(M + 1)T$ times.
\item The first sub diagonal of $H$ contains the entries $\gamma_i = g_ih_i,\ i = 1,2,
\ldots,N$ each repeated at least $(T - \theta)$ times. \een

Since $g_0$ and $\gamma_i$ are independent, we apply implication
$(3)$ of theorem \ref{thm:LB on DMT} to get
\[
d_H(r) \geq d_{H_d}(r) + d_{H_l}(r).
\]
Doing similar calculations are the same as in section
\ref{subsubsec:SAF guardtime no direct link} we get \bea d_H(r) &
\geq & \left(1 - \frac{(M + 1)(T + \theta)}{(M + 1)T}r\right) +
N\left(1 - \frac{(M + 1)(T + \theta)}{M(T - \theta)}r \right)\notag.
\eea As the number of slots $M$ tends to infinity we have
\[
d_H(r) \geq \left(1 - \frac{T + \theta}{T}r\right) + N\left(1 -
\frac{T + \theta}{T - \theta}r\right).
\] \epf

Thus the DMT $(N + 1)(1 - r)$ can be met with large $T$. This DMT
can be achieved by using a $(M + 1)T \times (M + 1)T$ approximately
universal CDA code.

\section{Slotted Amplify and Forward Protocol for the Slot-Offset Model \label{sec:SAF slot offset}}

\subsection{Slot-Offset Model}\label{subsec:Slot offset model description}

We propose another model for asynchronous cooperative relay
communication which we call the slot-offset model. In this model the
relay makes an error in detecting the beginning of its intended
listening epoch. For instance, in the SAF protocol the intended
listening epochs are slots, and the timing offset at a relay will
result in an offset in the beginning of the slot. This amounts to
slot-level asynchronism. The timing offset is assumed to be in units
of one symbol duration. We illustrate this model with an example.

Consider the two-hop network with a single source, single sink and
three relays. Let there be no direct link from the source to the
sink. We operate the SAF protocol with 3 slots, each of duration 3
channel uses, i.e., $M = 3$ and $T = 3$. Let the timing offset
profile be $[\tau_1 \ \tau_2 \ \tau_3] = [1 \ 0 \ 1]$. The source
transmits the vector
\[
\underline{x} = [x_0\ x_1\ x_2\ x_3\ x_4\ x_5\ x_6\ x_7\ x_8].
\]
As relay $R_1$ is offset by 1 channel use, it will listen to, and in
turn relay the symbols $[x_1\ x_2\ x_3]$ as opposed to relaying
$[x_0\ x_1\ x_2]$. Thus  it will miss the first symbol $x_0$.
Similarly relay $R_3$ will miss the first symbol in the third slot,
namely $x_6$ and relay the symbols $[x_7\ x_8\ 0]$ as opposed to
relaying $[x_6\ x_7\ x_8]$. Relay $R_2$, being perfectly
synchronized with the source, forwards the symbols that are assigned
to it, namely $[x_3\ x_4\ x_5]$. Notice that the symbol $x_3$ is
relayed by both $R_1$ and $R_2$ simultaneously, whereas $x_0$ is not
forwarded by any relay.

As evident from the above example we make the following observations
about the model:\ben \item Some symbols may not be relayed by any
relay node at all. For instance if the node which is supposed to
relay the first packet has a non-zero offset, then the first few
symbols will be lost. \item Some symbols may be listened and relayed
by more than one relays. Then the resulting schedule of receptions
will no longer be orthogonal as opposed to a possibly intended
orthogonal schedule.\een

Let $\tau_i$ denote the timing offset at relay $R_i$ and let $\theta
= \max_{1 \leq i \leq N} \tau_i$. We assume that there are no
relative propagation delays anywhere in the network. We further
assume that the destination is synchronized with the source
transmissions and it has perfect knowledge of timing offsets at all
the relays. This can be accomplished by a simple training scheme.

\subsection{Case with no Direct Link \label{subsec:Slot offset no direct link}}

First we consider the two hop network without the direct link and
all the relays connected. We operate the naive SAF protocol on this
network and analyze the impact of timing offsets on the DMT. The
start of the relay $R_i$ is offset by $\tau_i$ symbols as described
earlier. The length of one frame is taken to be $((M + 1)T +
\theta)$ channel uses, i.e., the source flushes $\theta$ zeros after
sending all the information symbols in order to isolate two
consecutive frames. We now describe the reception and transmission
time instants of the relays during the first cycle. This follows
periodically in the subsequent cycles.

Relay $R_1$ listens through the time instants $\tau_1 + 1$ to $T +
\tau_1$ and transmits the $T$ length vector it heard in the next
immediate $T$ channel uses. In general relay $R_i$ listens through
time instants $(i-1)T + \tau_N + 1 $ to $ iT + \tau_N $ and
transmits through time instants $iT + \tau_N + 1 $ to $ (i + 1)T +
\tau_N$.

The relay $R_i$ misses the first $\tau_i$ symbols in its respective
slot and receives the first $\tau_i$ symbols in the subsequent slot
corresponding to the relay $R_{i + 1}$. As a consequence of these
timing offsets the transmissions of different relays will not be
orthogonal and the same symbol might be relayed by more than one
relay. But it is easy to see that whenever the transmissions of two
relays overlap, the relays will always be sending the same symbol
simultaneously (probably with casual interference from the previous
information symbols). By the nature of the protocol at the most only
two relays can be transmitting simultaneously. Hence, in the
presence of timing offsets we will have terms like $(\gamma_i +
\gamma_{i+1})x$ at the destination, where $\gamma_i$'s are as
defined before.

The destination knows all the timing offsets and collects the vector
of received symbols. At the end of one frame the resulting channel
model can be written as, \bea
    \underline{y} & = & H\underline{x} + \underline{w} \label{eq:chanel model for SAF slot offset}
\eea where $\underline{x}$ is the vector of symbols transmitted by
the source, $H$ is a lower triangular matrix of size $MT \times MT$
and $\underline{w}$ corresponds to white noise. For the example
considered in section \ref{subsec:Slot offset model description},
the resulting channel model at the end of one frame will be, \bea Y
& = & \left[
\begin{array}{ccccccccc}
        0\\
        *& \gamma_1 \\
        *&*& \gamma_1 \\
        *&*&*& \beta_{12}  \\
        \vdots&&&*&\gamma_2 \\
        &&\vdots&&*&\gamma_2 \\
        &&&&\hdots&*&0\\
        &&&&&&*&\gamma_3\\
        *&&\hdots&&\hdots&&&*&\gamma_3
        \end{array}
        \right] \left[\begin{array}{c}x_0\\x_1\\x_2\\x_3\\x_4\\x_5\\x_6\\x_7\\x_8 \end{array}
        \right] + \underline{w}
\eea where $\beta_{12} = \gamma_1 + \gamma_2$.

\bthm \label{thm:LB slot offset model no DL} For the protocol
proposed in section \ref{subsec:Slot offset no direct link} the DMT
is lower bounded by \bea d(r) \geq N \left(1 - \frac{(M + 1)T +
\theta}{M(T - \theta)}r \right) \eea which tends to $N(1 - r)$ for
large $M$ and $T$. \ethm

\pf First we make some observations regarding the entries of the
matrix $H$ in equation (\ref{eq:chanel model for SAF slot offset})
and then go on to derive a lower bound on the DMT of the protocol.

The main diagonal of $H$ consists of: \ben
\item terms like $\gamma_i = h_ig_i$.
\item terms like ($\gamma_i + \gamma_{i + 1}$) which result due to the
same symbol getting relayed by two relays simultaneously. The number
of times $(\gamma_i + \gamma_{i + 1})$ appears is given by $(\tau_i
- \tau_{i + 1})^+$. See, for example, the matrix above. \een Let
$H_d$ be the diagonal matrix corresponding to $H$. Then,
\[
d_H(r) \geq d_{H_d}(r).
\]
We further lower bound $d_{H_d}(r)$ by dropping all the terms like
$(\gamma_i + \gamma_{i + 1})$ from the diagonal matrix since
dropping terms only decreases the mutual information. The exact
number of times for which these terms appear depends on the timing
offsets $\tau_i$. Instead of going for an exact count of the number
of individual terms $\gamma_i$, we lower bound it as follows. The
transmission of relay $R_i$ can interfere with either that of $R_{i
- 1}$ or that of $R_{i + 1}$. If $T$ is the slot length, the number
of times the term $\gamma_i$ appears in the main diagonal is given
by,
\[
    K(T - (\tau_{i - 1} - \tau_{i})^+ - (\tau_i - \tau_{i + 1})^+)
\]
which can be further lower bounded by $K(T - \theta)$. Therefore the
corresponding DMT can be computed as follows:\bea
Pr\{\log(\prod_{i=1}^{N}(1 + \rho|\gamma_i|^2)^{K(T - \theta)}) \leq
r^{'}\log\rho\} & = & Pr\{K(T - \theta)
\sum_{i=1}^{N}(1-\alpha_i)^{+} \leq r^{'}\} \eea where $|\gamma_i|^2
= \rho^{-\alpha_i}, i = 1,2,\ldots, N$ and $r^{'} = ((M + 1)T +
\theta)r$. We then have, \bea d_H(r) \geq N \left(1 - \frac{(M + 1)T
+ \theta}{M(T - \theta)}r \right)
 \eea which as $M$ tends to
infinity achieves $ N \left(1 - \frac{T}{T - \theta}r \right)$. \epf

Thus the above SAF protocol is tolerant to timing offsets and meets
the transmit diversity bound for large $T$. Even in this case we see
that the effect of asynchronism on DMT can be reduced by operating
the protocol over larger length. The achievability is again shown by
using approximately universal codes from CDA.

\subsection{Case with Direct Link \label{sec:slot offset direct link}}

Here we consider the case with a direct link from the source to the
destination under relay isolation. The source keeps transmitting
continuously over all the $(M + 1)$ slots. The relay strategy is the
same as above. The length of one frame is taken as $((M + 1)T +
\theta)$ channel uses. At the end of one frame, the destination
collects the vector $\underline{y}$ of received symbols. The channel
model is
\[
\underline{y} =  H\underline{x} + \underline{w}
\]
 where $\underline{x}$ is the input vector of
 length $(M + 1)T$ and $H$ is a lower triangular matrix of dimension
 $(M + 1)T \times (M + 1)T$. In fact $H$ is a banded matrix with
 just one sub-diagonal below the main diagonal and zeros elsewhere
 (as a result of the fact that all the relays are isolated). The main
 diagonal of $H$ consists of $g_0$ repeated $(M + 1)T$ times. The sub-diagonal
 of $H$ consists of each of the terms $\gamma_i$ repeated at least $K(T - \theta)$
 times (using the same argument as in the previous sub-section).
 Since $\gamma_i$'s and $g_0$ are independent, we can use implication
 $3$ of theorem \ref{thm:LB on DMT} to lower bound the DMT of the
 $H$ matrix. We have \bea
 d_H(r) &\geq& d_{H_d}(r) + d_{H_l}(r)\notag\\
&=& \left(1 - \frac{(M + 1)T + \theta}{(M + 1)T}r \right) + N
\left(1 - \frac{(M + 1)T + \theta}{M(T - \theta)}r \right) \eea
thereby asymptotically achieving the transmit diversity bound (for
large $M$ and $T$).

\section{DMT Optimal Codes \label{sec:DMT optimal codes}}

    As we had mentioned earlier, the distributed space time codes for
the above protocols are derived from Cyclic Division Algebras, see
\cite{SetRajSas,EliKumPawKumLu}. These codes are approximately
universal, see \cite{TavVis,EliKumPawKumLu} and achieve the DMT of
channels with arbitrary fading distribution. The resulting channel
model for the protocols we considered is given by,
\[
\underline{y} =  H\underline{x} + \underline{w}
\]
where $H$ is the induced  channel matrix with random fading
coefficients. If $H$ is of size, say, $L \times MT$ the DMT of $H$
can be achieved by an approximately universal CDA based code of size
$MT \times MT$. In the $k^{th}$ frame the source transmits the
$k^{th}$ column of the codeword matrix.

\section*{Acknowledgements}
Thanks are due to Birenjith, Sreeram and Vinodh for useful
discussions. This work is supported in part by NSF-ITR Grant
CCR-0326628 and in part by the DRDO-IISc Program on Advanced
Research in Mathematical Engineering.

\bibliographystyle{ieeetran}
\bibliographystyle{IEEEbib}

\end{document}